\pdfoutput=1
\documentclass[12pt,a4paper]{article}

\usepackage{ifthen} % for conditional statements
\newboolean{pdflatex}
\setboolean{pdflatex}{true} % False for eps figures 

\newboolean{articletitles}
\setboolean{articletitles}{true} % False removes titles in references

\newboolean{uprightparticles}
\setboolean{uprightparticles}{false} %True for upright particle symbols

\def\paperauthors{LHCb collaboration} % Leave as is for PAPER, CONF and FIGURE
\def\paperasciititle{Exotic hadron naming convention} % Set ASCII title here !! MAKE sure it's only ASCII characters !! 
\def\papertitle{Exotic hadron naming convention} % Latex formatted title
\def\paperkeywords{{High Energy Physics}, {LHCb}} % Comma separated list
\def\papercopyright{\the\year\ CERN for the benefit of the LHCb collaboration} % new since 9/Apr/2018
\def\paperlicence{CC BY 4.0 licence}
\def\paperlicenceurl{https://creativecommons.org/licenses/by/4.0/}

\usepackage[top=1in, bottom=1.25in, left=1in, right=1in]{geometry}

\usepackage{microtype}
\usepackage{lineno}  % for line numbering during review
\usepackage{xspace} % To avoid problems with missing or double spaces after
\usepackage{caption} %these three command get the figure and table captions automatically small

\usepackage{graphicx}  % to include figures (can also use other packages)
\usepackage{color}
\usepackage{colortbl}
\graphicspath{{./figs/}} % Make Latex search fig subdir for figures
\usepackage{amsmath} % Adds a large collection of math symbols
\usepackage{amssymb}
\usepackage{amsfonts}
\usepackage{upgreek} % Adds in support for greek letters in roman typeset

\newcommand*\patchAmsMathEnvironmentForLineno[1]{%
\expandafter\let\csname old#1\expandafter\endcsname\csname #1\endcsname
\expandafter\let\csname oldend#1\expandafter\endcsname\csname
end#1\endcsname
 \renewenvironment{#1}%
   {\linenomath\csname old#1\endcsname}%
   {\csname oldend#1\endcsname\endlinenomath}%
}
\newcommand*\patchBothAmsMathEnvironmentsForLineno[1]{%
  \patchAmsMathEnvironmentForLineno{#1}%
  \patchAmsMathEnvironmentForLineno{#1*}%
}
\AtBeginDocument{%
\patchBothAmsMathEnvironmentsForLineno{equation}%
\patchBothAmsMathEnvironmentsForLineno{align}%
\patchBothAmsMathEnvironmentsForLineno{flalign}%
\patchBothAmsMathEnvironmentsForLineno{alignat}%
\patchBothAmsMathEnvironmentsForLineno{gather}%
\patchBothAmsMathEnvironmentsForLineno{multline}%
\patchBothAmsMathEnvironmentsForLineno{eqnarray}%
}

\usepackage{hyperxmp}

\usepackage[pdftex,
            pdfauthor={\paperauthors},
            pdftitle={\paperasciititle},
            pdfkeywords={\paperkeywords},
            pdfcopyright={Copyright (C) \papercopyright},
            pdflicenseurl={\paperlicenceurl}]{hyperref}
\usepackage[colorinlistoftodos,textsize=scriptsize]{todonotes}

\usepackage[bottom,flushmargin,hang,multiple]{footmisc}

\usepackage[all]{hypcap} % Internal hyperlinks to floats.
\usepackage{comment}

\usepackage{xspace} 
\usepackage{upgreek}

\def\MagUp {\mbox{\em Mag\kern -0.05em Up}\xspace}

\ifthenelse{\boolean{uprightparticles}}%
{

 \def\Ppi         {\ensuremath{\uppi}\xspace}

 \def\Ppsi        {\ensuremath{\uppsi}\xspace}

 \def\PDelta      {\ensuremath{\Delta}\xspace}                 
 \def\PXi         {\ensuremath{\Xi}\xspace}                 
 \def\PLambda     {\ensuremath{\Lambda}\xspace}                 
 \def\PSigma      {\ensuremath{\Sigma}\xspace}                 
 \def\POmega      {\ensuremath{\Omega}\xspace}                 
 \def\PUpsilon    {\ensuremath{\Upsilon}\xspace}

 \def\PB      {\ensuremath{\mathrm{B}}\xspace}                 
                  
 \def\PD      {\ensuremath{\mathrm{D}}\xspace}

 \def\PJ      {\ensuremath{\mathrm{J}}\xspace}                 
 \def\PK      {\ensuremath{\mathrm{K}}\xspace}

 \def\Pb      {\ensuremath{\mathrm{b}}\xspace}                 
 \def\Pc      {\ensuremath{\mathrm{c}}\xspace}

 \def\Pp      {\ensuremath{\mathrm{p}}\xspace}

 \def\Ps      {\ensuremath{\mathrm{s}}\xspace}

 \def\thebaroffset{0.0em}
}
{

 \def\Ppi         {\ensuremath{\pi}\xspace}

 \def\Ppsi        {\ensuremath{\psi}\xspace}                 
                  
 \mathchardef\PDelta="7101
 \mathchardef\PXi="7104
 \mathchardef\PPi="7105
 \mathchardef\PLambda="7103
 \mathchardef\PSigma="7106
 \mathchardef\POmega="710A
 \mathchardef\PUpsilon="7107
                  
 \def\PB      {\ensuremath{B}\xspace}                 
                  
 \def\PD      {\ensuremath{D}\xspace}

 \def\PJ      {\ensuremath{J}\xspace}                 
 \def\PK      {\ensuremath{K}\xspace}

 \def\Pb      {\ensuremath{b}\xspace}                 
 \def\Pc      {\ensuremath{c}\xspace}

 \def\Pp      {\ensuremath{p}\xspace}

 \def\Ps      {\ensuremath{s}\xspace}

 \def\thebaroffset{0.18em}
}
\newcommand{\offsetoverline}[2][\thebaroffset]{\kern #1\overline{\kern -#1 #2}}%

\makeatletter
\ifcase \@ptsize \relax% 10pt
  \newcommand{\miniscule}{\@setfontsize\miniscule{4}{5}}% \tiny: 5/6
\or% 11pt
  \newcommand{\miniscule}{\@setfontsize\miniscule{5}{6}}% \tiny: 6/7
\or% 12pt
  \newcommand{\miniscule}{\@setfontsize\miniscule{5}{6}}% \tiny: 6/7
\fi
\makeatother

\DeclareRobustCommand{\optbar}[1]{\shortstack{{\miniscule (\rule[.5ex]{1.25em}{.18mm})}
  \\ [-.7ex] $#1$}}

\def\squark    {{\ensuremath{\Ps}}\xspace}

\def\cquark    {{\ensuremath{\Pc}}\xspace}

\def\bquark    {{\ensuremath{\Pb}}\xspace}

\def\pion   {{\ensuremath{\Ppi}}\xspace}

\def\pip    {{\ensuremath{\pion^+}}\xspace}
\def\pim    {{\ensuremath{\pion^-}}\xspace}

\def\kaon    {{\ensuremath{\PK}}\xspace}
\def\Kbar    {{\ensuremath{\offsetoverline{\PK}}}\xspace}
\def\Kb      {{\ensuremath{\Kbar}}\xspace}
\def\KorKbar {\kern \thebaroffset\optbar{\kern -\thebaroffset \PK}{}\xspace}
\def\Kz      {{\ensuremath{\kaon^0}}\xspace}
\def\Kzb     {{\ensuremath{\Kbar{}^0}}\xspace}
\def\Kp      {{\ensuremath{\kaon^+}}\xspace}
\def\Km      {{\ensuremath{\kaon^-}}\xspace}

\def\Dbar    {{\ensuremath{\offsetoverline{\PD}}}\xspace}
\def\D       {{\ensuremath{\PD}}\xspace}

\def\DorDbar {\kern \thebaroffset\optbar{\kern -\thebaroffset \PD}\xspace}
\def\Dz      {{\ensuremath{\D^0}}\xspace}
\def\Dzb     {{\ensuremath{\Dbar{}^0}}\xspace}
\def\Dp      {{\ensuremath{\D^+}}\xspace}
\def\Dm      {{\ensuremath{\D^-}}\xspace}

\def\DpDm    {\ensuremath{\Dp {\kern -0.16em \Dm}}\xspace}

\def\Dstarp  {{\ensuremath{\D^{*+}}}\xspace}
\def\Dstarm  {{\ensuremath{\D^{*-}}}\xspace}

\def\Ds      {{\ensuremath{\D^+_\squark}}\xspace}
\def\Dsp     {{\ensuremath{\D^+_\squark}}\xspace}
\def\Dsm     {{\ensuremath{\D^-_\squark}}\xspace}

\def\B       {{\ensuremath{\PB}}\xspace}
\def\Bbar    {{\ensuremath{\offsetoverline{\PB}}}\xspace}
\def\Bb      {{\ensuremath{\Bbar}}\xspace}
\def\BorBbar {\kern \thebaroffset\optbar{\kern -\thebaroffset \PB}\xspace}
\def\Bz      {{\ensuremath{\B^0}}\xspace}
\def\Bzb     {{\ensuremath{\Bbar{}^0}}\xspace}
\def\Bd      {{\ensuremath{\B^0}}\xspace}

\def\BdorBdbar {\kern \thebaroffset\optbar{\kern -\thebaroffset \Bd}\xspace}
\def\Bu      {{\ensuremath{\B^+}}\xspace}
\def\Bub     {{\ensuremath{\B^-}}\xspace}
\def\Bp      {{\ensuremath{\Bu}}\xspace}
\def\Bm      {{\ensuremath{\Bub}}\xspace}

\def\Bs      {{\ensuremath{\B^0_\squark}}\xspace}
\def\Bsb     {{\ensuremath{\Bbar{}^0_\squark}}\xspace}
\def\BsorBsbar {\kern \thebaroffset\optbar{\kern -\thebaroffset \Bs}\xspace}
\def\Bc      {{\ensuremath{\B_\cquark^+}}\xspace}

\def\Bcm     {{\ensuremath{\B_\cquark^-}}\xspace}

\def\jpsi     {{\ensuremath{{\PJ\mskip -3mu/\mskip -2mu\Ppsi}}}\xspace}

\def\Upsilonres  {{\ensuremath{\PUpsilon}}\xspace}
\def\Y#1S{\ensuremath{\PUpsilon{(#1S)}}\xspace}

\def\proton      {{\ensuremath{\Pp}}\xspace}

\def\Lz          {{\ensuremath{\PLambda}}\xspace}

\def\LorLbar     {\kern \thebaroffset\optbar{\kern -\thebaroffset \PLambda}\xspace}

\def\Sigmares    {{\ensuremath{\PSigma}}\xspace}

\def\Xires       {{\ensuremath{\PXi}}\xspace}

\def\Lc          {{\ensuremath{\Lz^+_\cquark}}\xspace}

\def\Xicz        {{\ensuremath{\Xires^0_\cquark}}\xspace}
\def\Xicp        {{\ensuremath{\Xires^+_\cquark}}\xspace}

\def\Lb           {{\ensuremath{\Lz^0_\bquark}}\xspace}

\newcommand{\decay}[2]{\ensuremath{#1\!\to #2}\xspace} 

\def\to                 {\ensuremath{\rightarrow}\xspace}

\def\AT#1     {\ensuremath{A_{\mathrm{T}}^{#1}}\xspace}           % 2

\def\C#1      {\ensuremath{\mathcal{C}_{#1}}\xspace}                       % 9
\def\Cp#1     {\ensuremath{\mathcal{C}_{#1}^{'}}\xspace}                    % 7
\def\Ceff#1   {\ensuremath{\mathcal{C}_{#1}^{\mathrm{(eff)}}}\xspace}        % 9  
\def\Cpeff#1  {\ensuremath{\mathcal{C}_{#1}^{'\mathrm{(eff)}}}\xspace}       % 7
\def\Ope#1    {\ensuremath{\mathcal{O}_{#1}}\xspace}                       % 2
\def\Opep#1   {\ensuremath{\mathcal{O}_{#1}^{'}}\xspace}                    % 7

\newcommand{\aunit}[1]{\ensuremath{\text{\,#1}}}       
\newcommand{\tev}{\aunit{Te\kern -0.1em V}\xspace}
\newcommand{\gev}{\aunit{Ge\kern -0.1em V}\xspace}
\newcommand{\mev}{\aunit{Me\kern -0.1em V}\xspace}
\newcommand{\kev}{\aunit{ke\kern -0.1em V}\xspace}
\newcommand{\ev}{\aunit{e\kern -0.1em V}\xspace}
 
\newcommand{\mevc}{\ensuremath{\aunit{Me\kern -0.1em V\!/}c}\xspace}
\newcommand{\gevc}{\ensuremath{\aunit{Ge\kern -0.1em V\!/}c}\xspace}
\newcommand{\mevcc}{\ensuremath{\aunit{Me\kern -0.1em V\!/}c^2}\xspace}
\newcommand{\gevcc}{\ensuremath{\aunit{Ge\kern -0.1em V\!/}c^2}\xspace}
\def\gsim{{~\raise.15em\hbox{$>$}\kern-.85em
          \lower.35em\hbox{$\sim$}~}\xspace}
\def\lsim{{~\raise.15em\hbox{$<$}\kern-.85em
          \lower.35em\hbox{$\sim$}~}\xspace}

\def\tell1  {TELL1\xspace}
\def\ukl1   {UKL1\xspace}

\newcommand{\ie}{\mbox{\itshape i.e.}\xspace}

\usepackage{cite} % Allows for ranges in citations
\usepackage{mciteplus}
\usepackage{multirow}
\usepackage{color}
\definecolor{lightgrey}{rgb}{0.2,0.2,0.2}
\usepackage{colortbl}

\begin{document}

%%%%%%%%%%%%%%%%%%%%%%%%%
%%%%% Title     %%%%%%%%%
%%%%%%%%%%%%%%%%%%%%%%%%%
\renewcommand{\thefootnote}{\fnsymbol{footnote}}
\setcounter{footnote}{1}
\begin{titlepage}

% Header ---------------------------------------------------
\vspace*{-1.5cm}

\noindent
\begin{tabular*}{\linewidth}{lc@{\extracolsep{\fill}}r@{\extracolsep{0pt}}}
\ifthenelse{\boolean{pdflatex}}% Logo format choice
{\vspace*{-1.2cm}\mbox{\!\!\!\includegraphics[width=.14\textwidth]{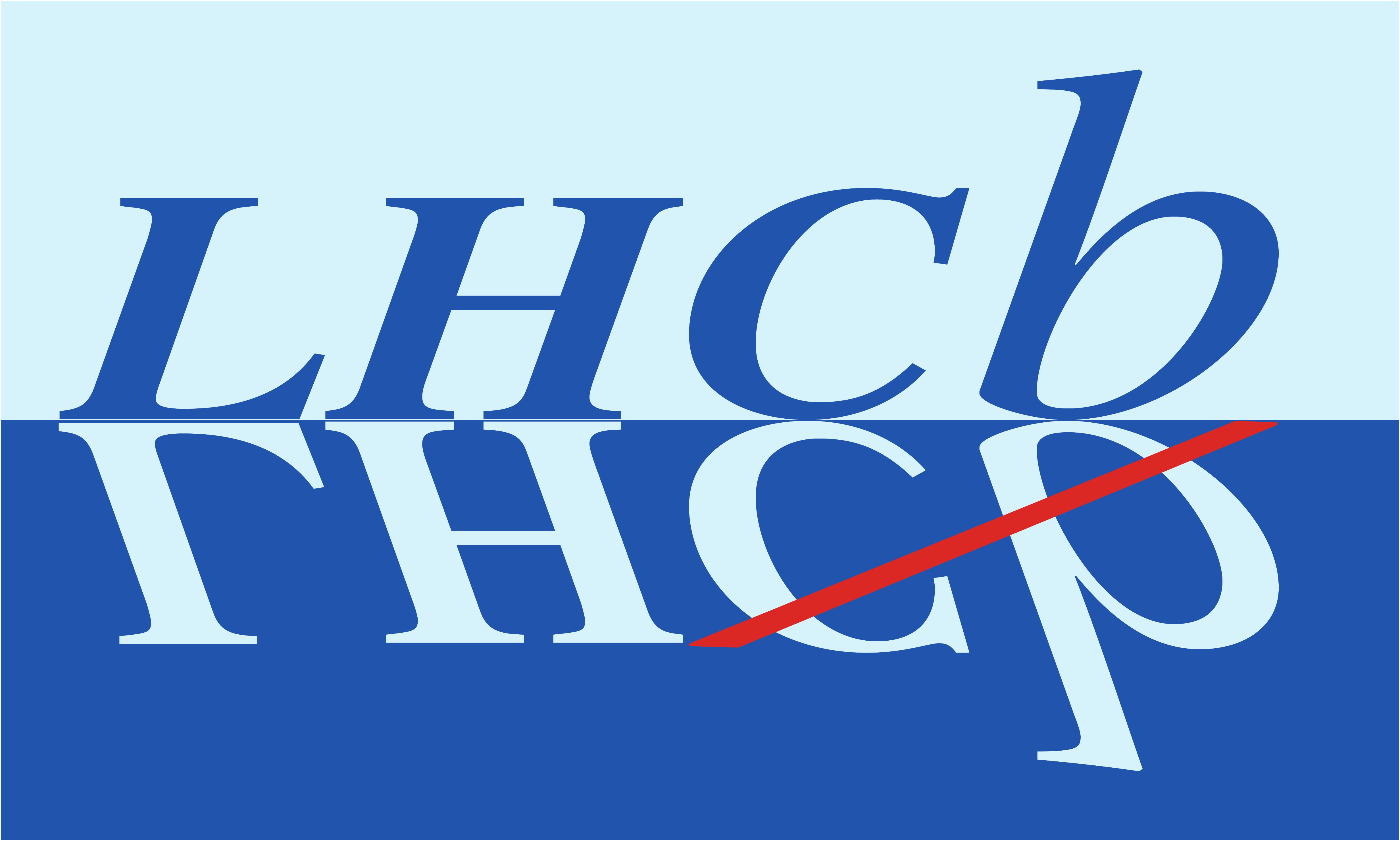}} & &}%
{\vspace*{-1.2cm}\mbox{\!\!\!\includegraphics[width=.12\textwidth]{figs/lhcb-logo.eps}} & &}
 \\
 & & LHCb-PUB-2022-013 \\  % ID 
 & & \today \\ % Date - Can also hardwire e.g.: 23 March 2010
 & & \\
\hline
\end{tabular*}

\vspace*{4.0cm}

% Title --------------------------------------------------
{\normalfont\bfseries\boldmath\huge
\begin{center}
% DO NOT EDIT HERE. Instead edit macro in main.tex to keep metadata correct
  \papertitle
\end{center}
}

\vspace*{2.0cm}

% Authors -------------------------------------------------
\begin{center}
  \paperauthors
  \footnote{Contact author: Tim Gershon, \href{mailto:T.J.Gershon@warwick.ac.uk}{T.J.Gershon@warwick.ac.uk}}
\end{center}

\vspace{\fill}

% Abstract -----------------------------------------------
\begin{abstract}
  \noindent 
  Many new exotic hadrons, that do not fit into the existing naming scheme for hadrons, have been discovered over the past few years.
  A new scheme is set out, extending the existing protocol, in order to provide a consistent naming convention for these newly discovered states, and other new hadrons that may be discovered in future.
\end{abstract}

\vspace*{2.0cm}
\vspace{\fill}
{\footnotesize
% Edit macro in main.tex to keep metadata correct
\centerline{\copyright~\papercopyright. \href{\paperlicenceurl}{\paperlicence}.}}
\vspace*{2mm}

\end{titlepage}
\pagestyle{empty}  % no page number for the title 
\newpage
\setcounter{page}{2}
\mbox{~}

\renewcommand{\thefootnote}{\arabic{footnote}}
\setcounter{footnote}{0}
\cleardoublepage

%%%%%%%%%%%%%%%%%%%%%%%%%
%%%%% Main text %%%%%%%%%
%%%%%%%%%%%%%%%%%%%%%%%%%

\pagestyle{plain} % restore page numbers for the main text
\setcounter{page}{1}
\pagenumbering{arabic}

\section{Introduction}
\label{sec:introduction}
 
\begin{figure}[!tb]
  \centering   
  \includegraphics[width=\textwidth]{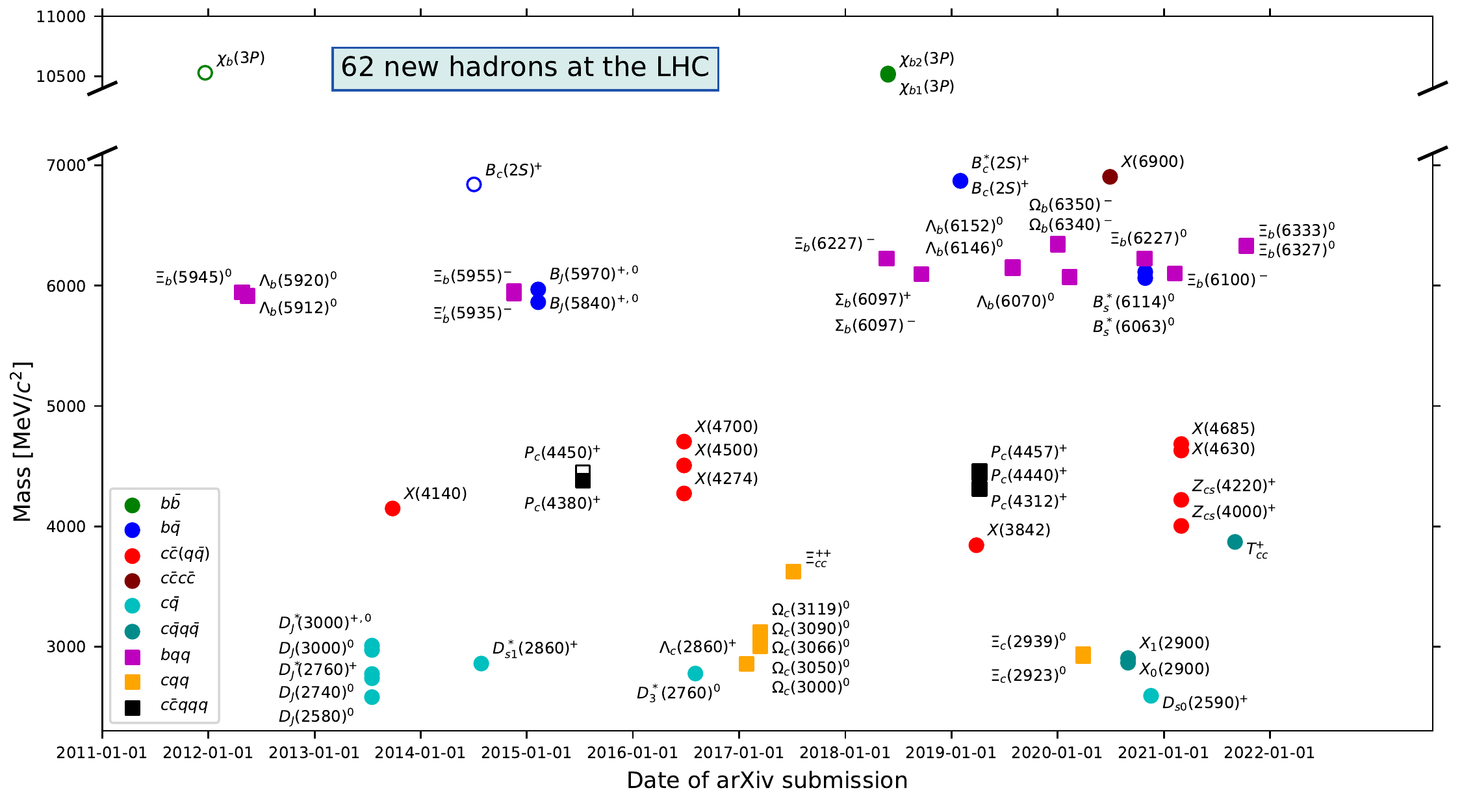}   
  \caption{
    Hadrons discovered at the LHC, plotted as mass versus preprint submission date~\cite{LHCb-FIGURE-2021-001}. 
    Only states observed with significance exceeding $5\sigma$ are included.
    Hollow markers indicate superseded states.
  }
  \label{fig:masses} 
\end{figure}

A large number of hadrons have been discovered using data collected by the LHC experiments, as summarised in Fig.~\ref{fig:masses} (taken from Ref.~\cite{LHCb-FIGURE-2021-001}).
These include the following states, which are considered manifestly exotic as their flavour quantum numbers do not fit into the scheme of conventional $q\bar{q}^\prime$ mesons or $qq^{\prime}q^{\prime\prime}$ baryons (where $q^{(\prime)(\prime)}$ indicates a quark of particular flavour).
\begin{itemize}
\item The charmonium-pentaquark $P_c(4380)^+$ and $P_c(4450)^+$ states~\cite{LHCb-PAPER-2015-029}, observed decaying to $\jpsi\proton$ in $\decay{\Lb}{\jpsi\proton\Km}$ decays.  The $P_c(4450)^+$ structure was subsequently resolved into two states, $P_c(4440)^+$ and $P_c(4457)^+$~\cite{LHCb-PAPER-2019-014}.
\item An additional narrow charmonium-pentaquark $P_c(4312)^+$ state~\cite{LHCb-PAPER-2019-014}, also seen decaying to $\jpsi\proton$ in $\decay{\Lb}{\jpsi\proton\Km}$ decays.
\item The $X(6900)$ state observed in the $\jpsi\jpsi$ final state~\cite{LHCb-PAPER-2020-011}.
\item The $X_0(2900)$ and $X_1(2900)$ states observed as $\Dm\Kp$ resonances in $\decay{\Bp}{\Dp\Dm\Kp}$ decays~\cite{LHCb-PAPER-2020-024,LHCb-PAPER-2020-025}.
\item The $Z_{cs}(4000)^+$ and $Z_{cs}(4220)^+$ states, seen decaying to $\jpsi\Kp$ in $\decay{\Bp}{\jpsi\phi\Kp}$ decays~\cite{LHCb-PAPER-2020-044}.
\item The $T_{cc}(3875)^+$ state observed in the $\Dz\Dz\pip$ final state~\cite{LHCb-PAPER-2021-031,LHCb-PAPER-2021-032}.
\end{itemize}
The above list does not include the charmonium-like $X$ states, which do not have manifestly exotic quantum numbers.
The LHC discoveries complement those that have been made at other experiments, including BESIII, Belle and BaBar.
Detailed reviews can be found in Refs.~\cite{Ali:2017jda,Olsen:2017bmm,Karliner:2017qhf,Guo:2017jvc,Liu:2019zoy,Brambilla:2019esw}.

The current naming scheme, set out in the Particle Data Group's Review of Particle Physics~\cite{PDG2020}, does not fully cover states with manifestly exotic quantum numbers.
As such, several of the names that have been assigned do not follow any defined scheme.
It is therefore well motivated to develop a convention that can be used for these and any future discoveries.
This may hopefully minimise the amount of confusion caused by the diverse range of states that are possible.  
Indeed, the existing naming scheme arose from similar concerns about the ever expanding ``particle zoo'' of hadrons that were discovered in the 1960s, 70s and 80s~\cite{CERNCourierVolume25:1731195}.

\section{Current naming scheme for hadrons}
\label{sec:PDGscheme}
 
The current naming conventions are set out in the Particle Data Group's Review of Particle Physics~\cite{RPP-naming,PDG2020}, and are broadly accepted in the community.
These are briefly reviewed here, as any exotic hadron naming scheme should build on this existing framework.
A key point is that the naming is based on measured quantum numbers, rather than speculation about the degrees of freedom within the hadron (such as whether a molecular-like binding mechanism is at work, or whether a state has gluonium content).\footnote{
  Exceptions are made for states with clearly established spectroscopic identities, for example $\psi(2S)$.
}
Table~\ref{tab:PDG-mesons} summarises the convention for mesons with zero net strangeness and heavy flavour quantum numbers.
These symbols are supplemented by a subscript $J$ to indicate the spin, except for pseudoscalar and vector mesons.
The mass, in units of \mevcc, is added in parentheses after the symbol for mesons that decay strongly, though in practice the mass is often omitted for familiar states.
The electric charge is also specified as a superscript, as appropriate, \ie\ it can be omitted for isospin singlet ($I=0$) states where only one charge is possible.

\begin{table}[!htb]
  \centering
  \renewcommand*{\arraystretch}{1.1}
  \caption{
    Naming convention for mesons with zero net strangeness and heavy flavour quantum numbers, from Ref.~\cite{PDG2020}.
    The indicated spin $J$ for given parity $P$ and charge conjugation $C$ quantum numbers, is the lowest possible in the naive quark model.
    The symbols, however, can be used for any spin, which is indicated by a subscript.
    The two $I=0$ states mix, and the mass eigenstates use the same symbol, distinguished by the addition of a prime ($^\prime$) for the heavier, except for the vector ($J^{PC} = 1^{--}$) states where near-maximal flavour $SU(3)$-breaking justifies the use of different symbols for the two states.  
}
  \label{tab:PDG-mesons}
  \begin{tabular}{lcccc}
    \hline
    \multicolumn{1}{r}{$J^{PC}\ ^{(1)}$} & $0^{-+}$ & $1^{+-}$ & $1^{--}$ & $0^{++}$ \\
%                                 & $2^{-+}$ & $3^{+-}$ & $2^{--}$ & $1^{++} \\
    Minimal quark content & \\
    \hline
    $u\bar{d}$, $u\bar{u}-d\bar{d}$, $\bar{u}d$ ($I=1$) & $\pi$ & $b$ & $\rho$ & $a$ \\
    $u\bar{u}+d\bar{d}$ and/or $s\bar{s}$ ($I=0$) & $\eta^{(\prime)}$ & $h^{(\prime)}$ & $\omega$, $\phi$ & $f^{(\prime)}$ \\
    $c\bar{c}$ & $\eta_c$ & $h_c$ & $\psi\ ^{(2)}$ & $\chi_c$ \\
    $b\bar{b}$ & $\eta_b$ & $h_b$ & $\Upsilonres$ & $\chi_b$ \\
   \hline
  \end{tabular}
%  \newline
  \begin{flushleft}
    {\scriptsize ${}^{(1)}$ The $C$ quantum number is only defined for the neutral member of an isospin triplet, but it can be extended to $G$-parity which is defined for all mesons in this table.
    As the $G$ eigenvalue is equal to $(-1)^I$ times the $C$ eigenvalue, one has $G = -$ for $\pi, h^{(\prime)}, \omega/\phi, a$ and $G = +$ for $\eta^{(\prime)}, b, \rho, f^{(\prime)}$.
    } \\
    {\scriptsize ${}^{(2)}$ The $\jpsi$ state retains its historical dual name, as an exception to this scheme.}
  \end{flushleft}
\end{table}

Following the discovery of charmonium-like and bottomonium-like states with $I=1$, this scheme has been expanded to introduce the symbols $Z_c$ and $Z_b$ for such states with $PC = +-$ (see note ${}^{(1)}$ in Table~\ref{tab:PDG-mesons}).
The symbols $\PPi_{c,b}$, $R_{c,b}$ and $W_{c,b}$ have likewise been reserved for such states with $PC$ = $-+$, $--$ and $++$ respectively, although few such states are currently known.
The symbol $X$ is reserved for states with unknown quantum numbers.

The naming convention for $q\bar{q}^\prime$ mesons with non-zero strangeness, charm or beauty quantum numbers is given in Table~\ref{tab:HF-mesons}.
As for mesons with zero net strangeness or heavy flavour quantum numbers, the symbols are supplemented by a subscript $J$ to indicate the spin, though this is usually omitted for the lightest pseudoscalar and vector mesons of given flavour quantum numbers.
The mass, in units of \mevcc, is added in parentheses for mesons that decay strongly, and a superscript $*$ is added for states with natural spin-parity, \ie\ $J^P$ in the series $0^+, 1^-, 2^+$.
For these states, the electric charge is always specified as a superscript.
The charge-conjugate neutral mesons are distinguished by an overline, following the convention $s \to  \Kb$, $c \to D$, $b \to \Bb$.
As a result, and with the convention that the flavour quantum number and the charge of a quark have the same sign, the $\Kz$, $\Dz$ and $\Bz$ mesons have positive strangeness, charm and beauty quantum numbers, respectively.  

\begin{table}[!htb]
  \centering
  \renewcommand*{\arraystretch}{1.15}
  \caption{
    Naming convention for $q\bar{q}^\prime$ mesons with non-zero strangeness, charm or beauty quantum numbers, from Ref.~\cite{PDG2020}.
  }
  \label{tab:HF-mesons}
  \begin{tabular}{|l|ccccc|}
    \hline % \\ [-2.3ex]
    & $\bar{u}$ & $\bar{d}$ & $\bar{s}$ & $\bar{c}$ & $\bar{b}$ \\
    \hline
    $u$ & \multicolumn{2}{c}{\cellcolor{lightgray}} & \Kp & \Dzb & \Bp \\
    $d$ & \multicolumn{2}{c}{\multirow{-2}{*}{\cellcolor{lightgray}{\small See Table~\ref{tab:PDG-mesons}}}} & \Kz & \Dm & \Bz \\
    $s$ & \Km & \Kzb & \cellcolor{lightgray} & \Dsm & \Bs \\
    $c$ & \Dz & \Dp & \Ds & \cellcolor{lightgray} & \Bc \\
    $b$ & \Bm & \Bzb & \Bsb & \Bcm & \cellcolor{lightgray} \\
    \hline
  \end{tabular}
\end{table}

The naming convention for baryons, summarised in Table~\ref{tab:baryons}, consists of a symbol that identifies the isospin, plus subscripts that identify the heavy flavour quantum numbers.  
The symbol also distinguishes between the $I=\frac{1}{2}$ cases where all three of the constituent quarks are $u$ or $d$, and that where only one is; a similar distinction is made for $I=0$ with two or zero $u$ or $d$ quarks.
The additional quark(s) are implicitly strange unless $c$ or $b$ subscripts are given.
The mass, in units of \mevcc, is given in parentheses if the baryon decays strongly.
Furthermore the spin-parity $J^P$ quantum numbers should be specified after the name, although in practice these are often omitted.  
A $*$ superscript is sometimes added to indicate an excited state, although this is not specified in the convention~\cite{RPP-naming,RPP-quarkmodel}.
The electric charge is also specified as a superscript, as appropriate (\ie\ it can be omitted for $I=0$ states where only one charge is possible).
Baryons are always written without, and antibaryons always with, an overline.

\begin{table}[!htb]
  \centering
  \renewcommand*{\arraystretch}{1.1}
  \caption{
    Naming convention for baryons, from Ref.~\cite{PDG2020}.
    There are two $J^P = \frac{1}{2}^+$ $\PXi_c$ and $\PXi_b$ isospin doublets in the respective flavour $SU(4)$ 20-plets; these are distinguished by the addition of a prime to the symbol for the heavier~\cite{RPP-quarkmodel}.
  }
  \label{tab:baryons}

\begin{tabular}{c@{\hspace{1cm}}c@{\hspace{1cm}}c}
  Three $u/d$ quarks & Two $u/d$ quarks & One or zero $u/d$ quarks \\
  \begin{tabular}{cc}
    \hline
    $I=\frac{1}{2}$ & $I=\frac{3}{2}$ \\
    $N$ & $\PDelta$ \\
    \hline
    & \\
    & \\
    & \\
    & \\
  \end{tabular}
  & 
  \begin{tabular}{cc}
    \hline
    $I=0$ & $I=1$ \\
    $\PLambda$ & $\PSigma$ \\
    $\PLambda_c$ & $\PSigma_c$ \\
    $\PLambda_b$ & $\PSigma_b$ \\
    \hline
    & \\
    & \\
  \end{tabular}
  &
  \begin{tabular}{cc}
    \hline
    $I=\frac{1}{2}$ & $I=0$ \\
    $\PXi$ & $\POmega$ \\
    $\PXi_c^{(\prime)}$ & $\POmega_c$ \\
    $\PXi_b^{(\prime)}$ & $\POmega_b$ \\
    $\PXi_{cc}$ & $\POmega_{cc}$ \\
    \ldots & \ldots \\
    \hline
  \end{tabular}
\end{tabular}

\end{table}

\section{Shortcomings of the current naming scheme}
\label{sec:shortcomings}

The need for an extension of the naming scheme can be illustrated by considering the symbols assigned to the recently discovered, manifestly exotic, hadrons discussed in Sec.~\ref{sec:introduction}.
For this purpose, it is also useful to consider some states for which evidence, not at the $5\sigma$ discovery level, has been found.

\begin{itemize}
\item The naming scheme summarised in Sec.~\ref{sec:PDGscheme} does not provide for charmonium-pentaquark states, and the notation of the discovery paper~\cite{LHCb-PAPER-2015-029}, $P_c$, has been adopted by the community~\cite{RPP-naming}. 
The observed $P_c$ states have $I=\frac{1}{2}$. 
It is not obvious what symbol should be assigned to $I=\frac{3}{2}$ charmonium-pentaquark states, if they were to be discovered.
\item Possible $\jpsi \Lz$ resonances have been denoted $P_{cs}$~\cite{LHCb-PAPER-2020-039}.  Note that here the subscript $c$ indicates that the pentaquark has $c\bar{c}$ (hidden charm) content, while the $s$ subscript indicates that it contains an $s$ quark (open strangeness).  
These states have $I=0$, and if $I=1$ counterparts ($\jpsi \Sigmares$ resonances) are discovered it is not clear what symbol should be assigned to them.
\item In the event of a discovery of a pentaquark state with open charm, there is no obvious logical symbol to use.
\item The quantum numbers of the $X_0(2900)$ and $X_1(2900)$ states are all measured, except for isospin.  This has deferred the problem of what to name them, as the current naming scheme does not provide for a state with minimal quark content of $cs\bar{u}\bar{d}$.
\item The $Z_{cs}(4000)^+$ and $Z_{cs}(4220)^+$ states have isospin $I=\frac{1}{2}$, although the $Z$ symbol had prior to their discovery been reserved for states with $I=1$.  
Note that again the $c$ and $s$ subscripts indicate hidden charm and open strangeness, respectively.
\end{itemize}

\section{Exotic hadron naming scheme}
\label{sec:proposal}

Ideally it would be possible to make a simple and logical naming convention, that conveys in a compact way the quantum numbers (isospin, strangeness, heavy flavour, spin, parity and $G$-parity) of all possible hadrons.
However, the current naming scheme for hadrons, summarised in Sec.~\ref{sec:PDGscheme}, is already quite complex.
Any system that builds on it must be at least equally elaborate.
Nonetheless, it is important to retain backwards-compatibility, as much as possible, as doing otherwise will inevitably cause confusion.
Other desirable features of a naming scheme are to avoid conflicts with existing particle names (including those of leptons and bosons) and to be sufficiently flexible to accommodate a range of possible future discoveries.

With these points in mind, the following exotic hadron naming convention is proposed.
\begin{enumerate}
\item As in the current scheme, symbols are assigned based on measured quantum numbers, rather than speculation about the degrees of freedom within the hadron.
\item There is no change to the existing scheme for any state that does not unambiguously have four- or five-quark content.
\item States with minimum four-quark content are labelled $T$; states with minimum five-quark content are labelled $P$. $T$ states are mesons, while $P$ states are baryons.\footnote{
    The current scheme is limited to states with minimum four- or five-quark content.  
    It should, however, be possible to extend later to states with minimum six- or seven-quark content, if necessary.
    It is also limited to the allowed isospin values of conventional hadrons, although in principle value of $I$ up to $2$ for $T$ states and $\frac{5}{2}$ for $P$ states are possible.
    Extension of the scheme at a later date to cover more isospin values will also be possible if necessary.}
\item A superscript is added to indicate the isospin, and (where appropriate) parity and $G$-parity.  The superscripts follow existing conventions for labelling these properties, which differ for the cases of mesons with zero or non-zero net strangeness and heavy flavour quantum numbers, and differ again for baryons.
\begin{enumerate}
\item For $T$ states with zero net strangeness and heavy flavour quantum numbers, the symbols from the first two rows of Table~\ref{tab:PDG-mesons} are used, to cover the eight possibilities of $I = 0,1$, and even or odd $P$ and $G$-parity.  The symbol $\omega$ is used (instead of $\phi$) for the case $I=0$, $P$ and $G$ odd.
\item For $T$ states with non-zero strangeness, charm or beauty quantum numbers, the $G$-parity is not defined, and therefore the superscript encodes only the isospin and parity.  The labels $\eta, \tau, \pi$ are used for $I = 0, \frac{1}{2}, 1$ with $P$ odd, and $f, \theta, a$ are used for $I = 0, \frac{1}{2}, 1$ with $P$ even.\footnote{
    The symbols for $I = \frac{1}{2}$ refer to the historical ``$\tau$--$\theta$'' puzzle, in which decays of the kaon to three or two pions, which are respectively parity-odd and parity-even final states, were denoted by $\tau$ and $\theta$~\cite{Dalitz:1953cp,Lee:1956qn,Dalitz:1993yt}.}
\item For $P$ states, the superscript indicates the isospin only: it is $\PLambda, N, \PSigma, \PDelta$ for $I = 0, \frac{1}{2}, 1, \frac{3}{2}$.
\end{enumerate}
These superscripts are summarised in Table~\ref{tab:superscripts}. 
It should be stressed that the superscript does not convey the spin (see below); the distinction between $\pi$ and $\rho$ is thus $G$-parity rather than spin, as is also formally the case for conventional mesons in the PDG convention~\cite{RPP-naming}.
Note also that there is no need to add $*$ to denote natural spin-parity mesons.

\begin{table}[!htb]
  \centering
  \renewcommand*{\arraystretch}{1.1}
  \caption{
    Superscripts added to indicate isospin, and (where appropriate) parity and $G$-parity in the exotic hadron naming scheme.
  }
  \label{tab:superscripts}

\resizebox{\textwidth}{!}{
\begin{tabular}{c@{\hspace{5mm}}c@{\hspace{5mm}}c}
    $T$ states & $T$ states & $P$ states \\ 
    {\small zero net $S$, $C$, $B$} & {\small non-zero net $S$, $C$, $B$} & \\
    \begin{tabular}{ccc}
    \hline
    $(P,G)$ & $I=0$ & $I=1$ \\
    $(-,-)$ & $\omega$ & $\pi$\\
    $(-,+)$ & $\eta$ & $\rho$ \\
    $(+,+)$ & $f$ & $b$ \\
    $(+,-)$ & $h$ & $a$ \\
    \hline
    \end{tabular}
    & 
    \begin{tabular}{cccc}
    \hline
    $(P)$ & $I=0$ & $I = \frac{1}{2}$ & $I=1$ \\
    $(-)$ & $\eta$ & $\tau$ & $\pi$ \\
    $(+)$ & $f$ & $\theta$ & $a$ \\
    \hline
    \\
    \\
    \end{tabular}
    &
    \begin{tabular}{cccc}
    \hline   
    $I = 0$ & $I = \frac{1}{2}$ & $I = 1$ & $I = \frac{3}{2}$ \\
    $\PLambda$ & $N$ & $\PSigma$ & $\PDelta$ \\
    \hline
    \\
    \\
    \\
    \end{tabular}
\end{tabular}
}
\end{table}

\item Subscripts $\Upsilonres$, $\psi$ and $\phi$ are added to denote hidden beauty, charm and strangeness.  These should be in order of mass, where more than one is needed, and can be repeated if necessary.  Since $s\bar{s}$ content may mix with $u\bar{u}+d\bar{d}$ content, the $\phi$ subscript should only be applied where the $s\bar{s}$ content is clear.  It must be stressed that this reuse of the symbols for vector mesons conveys information only about the quark content, not about their arrangement or other quantum numbers.
\item Subscripts $b$, $c$ and $s$ are added to denote open flavour quantum numbers.  These should be in order of mass, where more than one is needed, and can be repeated if necessary.  They indicate the quark content, rather than the flavour quantum number.
If only one such symbol is needed, it should always be that of the quark ($b$, $c$, $s$) rather than the antiquark ($\bar{b}$, $\bar{c}$, $\bar{s}$).
If more than one is needed, the symbols can be those of quark or antiquark but the first should always be a quark.
If subscripts are needed to indicate both hidden and open flavour, the symbols for hidden flavour should appear first. 
\item For $T$ states an additional subscript should be added to indicate the spin $J$, while for $P$ states, the spin-parity should be added after the name.
These labels are thus added in the same way as in the current naming schemes for mesons and baryons, respectively.  This information is omitted when it is not yet known.
\item For $T$ states with open flavour quantum numbers, the particle and antiparticle are distinguished by the addition of an overline to the latter where necessary.
  This is needed for all such neutral $T$ states, and also for some charged $T$ states with two different open flavours (specifically, $T_{cs}$, $T_{bc}$, and $T_{b\bar{s}}$). 
  The ``particle'' is chosen to be that with the quark content specified in the subscript,\footnote{
    The choice of which meson is particle and which is antiparticle does not have any physical significance, it is simply whether the symbol does not or does include an overline.
} \ie\ where the heaviest open flavour is a quark rather than an antiquark.
  Overlines can also be added in cases where they are not strictly necessary if this is felt to add clarity.
  % Overlines are never used for charged $T$ states, as in the current scheme for mesons. 
$P$ states with baryon number $+1$ ($-1$) are always written without (with) an overline, as in the current scheme for baryons.

\item For all hadrons the mass, in units of \mevcc, should be added in parentheses, and the charge superscript should be added where appropriate.
\item In practice it is anticipated that some labels will be omitted where they are considered unnecessary.
\end{enumerate}

\section{Impact on discovered and hypothetical new states}

The impact on various recently discovered hadrons is summarised in Table~\ref{tab:summary}.
As stated previously, there is no change for any hadron with quantum numbers that are consistent with a conventional $q\bar{q}^\prime$ meson or $qq^{\prime}q^{\prime\prime}$ baryon.
For example, the $\chi_{c1}(3872)$ -- also known as $X(3872)$ -- is unchanged.
There is, however, a change for the $I=1$ charmonium-like and bottomonium-like states, currently denoted $Z_c$ and $Z_b$ --- these become $T^{b}_{\psi}$ and $T^{b}_{\Upsilonres}$ states, respectively.

\begin{table}[!tb]
  \renewcommand*{\arraystretch}{1.1}
  \centering
  \caption{
    Summary of the impact of the exotic hadron naming scheme on various states, based on current knowledge of their properties. 
    Quantum numbers that are not specified or marked ``?'' are unknown and the corresponding super-/sub-scripts not given.
    The current name indicated is that used in the PDG listings~\cite{PDG2020}.
  }
  \label{tab:summary}
  \begin{tabular}{ccccc}
    \hline
    {Minimal quark} & \multirow{2}{*}{Current name} & \multirow{2}{*}{$I^{(G)}$, $J^{P(C)}$} & \multirow{2}{*}{Proposed name} & \multirow{2}{*}{Reference} \\
    {content} \\
    \hline
    $c\bar{c}$ & $\chi_{c1}(3872)$ & $I^G=0^+$, $J^{PC} = 1^{++}$ & $\chi_{c1}(3872)$ & \cite{Belle:2003nnu,LHCb-PAPER-2013-001} \\
    $c\bar{c}u\bar{d}$ & $Z_c(3900)^+$ & $I^G=1^+$, $J^{P} = 1^{+}$ & $T^{b}_{\psi 1}(3900)^+$ & \cite{BESIII:2013ris,Belle:2013yex,BESIII:2017bua} \\
    $c\bar{c}u\bar{d}$ & $X(4100)^+$ & $I^G=1^-$ & $T_{\psi}(4100)^+$ & \cite{LHCb-PAPER-2018-034} \\
    $c\bar{c}u\bar{d}$ & $Z_c(4430)^+$ & $I^G=1^+$, $J^{P} = 1^{+}$ & $T^{b}_{\psi 1}(4430)^+$ & \cite{Belle:2007hrb,LHCb-PAPER-2014-014} \\ 
    $c\bar{c}(s\bar{s})$ & $\chi_{c1}(4140)$ & $I^G = 0^+, J^{PC} = 1^{++}$ & $\chi_{c1}(4140)$ & \cite{CDF:2009jgo,CMS:2013jru,LHCb-PAPER-2016-018,LHCb-PAPER-2016-019} \\
    $c\bar{c}u\bar{s}$ & $Z_{cs}(4000)^+$ & $I=\frac{1}{2}$, $J^P = 1^+$ & $T^\theta_{\psi s 1}(4000)^+$ & \cite{LHCb-PAPER-2020-044} \\
    $c\bar{c}u\bar{s}$ & $Z_{cs}(4220)^+$ & $I=\frac{1}{2}$, $J^P = 1^?$ & $T_{\psi s 1}(4220)^+$ & \cite{LHCb-PAPER-2020-044} \\
    $c\bar{c}c\bar{c}$ & $X(6900)$ & $I^G=0^+$, $J^{PC} = ?^{?+}$ & $T_{\psi\psi}(6900)$ & \cite{LHCb-PAPER-2020-011} \\
    $cs\bar{u}\bar{d}$ & $X_0(2900)$ & $J^P = 0^+$ & $T_{cs0}(2900)^0$ & \cite{LHCb-PAPER-2020-024,LHCb-PAPER-2020-025} \\
    $cs\bar{u}\bar{d}$ & $X_1(2900)$ & $J^P = 1^-$ & $T_{cs1}(2900)^0$ & \cite{LHCb-PAPER-2020-024,LHCb-PAPER-2020-025} \\
    $cc\bar{u}\bar{d}$ & $T_{cc}(3875)^+$ & & $T_{cc}(3875)^+$ & \cite{LHCb-PAPER-2021-031,LHCb-PAPER-2021-032} \\
    $b\bar{b}u\bar{d}$ & $Z_b(10610)^+$ & $I^G=1^+$, $J^{P} = 1^{+}$ & $T^{b}_{\Upsilonres 1}(10610)^+$ & \cite{Belle:2011aa} \\
    $c\bar{c}uud$ & $P_c(4312)^+$ & $I=\frac{1}{2}$ & $P^N_{\psi}(4312)^+$ & \cite{LHCb-PAPER-2019-014} \\
    $c\bar{c}uds$ & $P_{cs}(4459)^0$ & $I=0$ & $P^{\PLambda}_{\psi s}(4459)^0$ & \cite{LHCb-PAPER-2020-039} \\
    \hline
  \end{tabular}
\end{table}

The names specified in Table~\ref{tab:summary} are based on known quantum numbers.  
In several examples, only certain sets of quantum numbers are consistent with the measured properties. 
The following comments indicate how the name will change once these are experimentally established.
\begin{itemize}
\item The $J^P$ quantum numbers of the $X(4100)^+$ state are $0^+$ or $1^-$~\cite{LHCb-PAPER-2018-034}.  Once one of these is confirmed, the name will become $T^{a}_{\psi 0}(4100)^+$ or $T^{\pi}_{\psi 1}(4100)^+$, respectively.
\item The $\chi_{c1}(4140)$ --- also known as $X(4140)$ --- state is one of a number of resonances in the $\jpsi\phi$ system observed in $\Bp \to \jpsi\phi\Kp$ decays~\cite{CDF:2009jgo,CMS:2013jru,LHCb-PAPER-2016-018,LHCb-PAPER-2016-019,LHCb-PAPER-2020-044}.
It appears likely that it could have a predominant $c\bar{c}s\bar{s}$ component, but since this is not yet established experimentally its proposed name is unchanged from its current name.
In case a predominant $c\bar{c}s\bar{s}$ component is later established, for example through a stringent limit on its decay to $\jpsi\omega$, it can be accommodated in the new scheme as the $T^f_{\psi\phi1}(4140)$.
\item The $J^P$ quantum numbers of the $Z_{cs}(4220)^+$ state are $1^+$ or $1^-$~\cite{LHCb-PAPER-2020-044}.  Once one of these is confirmed, the name will become $T^\theta_{\psi s 1}(4220)^+$ or $T^\tau_{\psi s 1}(4220)^+$, respectively.
\item If a neutral isospin partner of the $Z_{cs}(4000)^+$ state is discovered, its name will be $T^\theta_{\psi s1}(4000)^0$ for the state containing $c\bar{c}s\bar{d}$.  Its antiparticle, containing $c\bar{c}\bar{s}d$ will be named $\overline{T}{}^\theta_{\psi s1}(4000)^0$. 
\item The $J^P$ quantum numbers of the $X(6900)$ state are yet to be determined.  Once known, the spin will be indicated as a subscript, and the parity as a superscript which will be either $\eta$ (odd) or $f$ (even).
\item Each of the $X_0(2900)$ and $X_1(2900)$ states could have either isospin $0$ or $1$.  In case of $I=0$ their names will become $T^{f}_{cs0}(2900)$ and $T^{\eta}_{cs1}(2900)$, while for $I=1$ their names will become $T^{a}_{cs0}(2900)^0$ and $T^{\pi}_{cs1}(2900)^0$, respectively.
The antiparticles of these neutral $T$ states are $\overline{T}{}_{cs0}(2900)$ and $\overline{T}{}_{cs1}(2900)$.
\item If the $X(2900)$ states have $I=1$, their charged partners will have particle and antiparticle distinguished by overlines.
  For example, the $J=0$ states with quark content $cs \bar{d}\bar{d}$ and $cs \bar{u}\bar{u}$ would be denoted $T^a_{cs0}(2900)^+$ and $T^a_{cs0}(2900)^-$, respectively, while their antiparticles with quark content $\bar{c}\bar{s} dd$ and $\bar{c}\bar{s} uu$ would be denoted $\overline{T}{}^a_{cs0}(2900)^-$ and $\overline{T}{}^a_{cs0}(2900)^+$, respectively.
\item The $T_{cc}(3875)^+$ state has properties consistent with expectations for $I=0$, $J^P = 1^+$~\cite{LHCb-PAPER-2021-031,LHCb-PAPER-2021-032}.  If those quantum numbers are experimentally established, the name will become $T^{f}_{cc1}(3875)^+$.
\end{itemize}

The main impact of the exotic hadron naming scheme is anticipated to be on states that may be discovered in future.
Any discussion of this is by definition speculative, but given the number of discoveries over the last few years it seems reasonable to expect that many more states can be found with existing and forthcoming data, especially considering that detector upgrades will provide better precision.
The existing scheme allows for a large number of hypothesised new hadrons, including doubly-heavy tetraquark ($T_{bc}$ and $T_{bb}$) states, isospin partners to the $D_{sJ}$ states~\cite{BaBar:2003oey,CLEO:2003ggt}, and a range of possible pentaquark states including those with open beauty or charm quantum numbers, or both.
Some examples of exotic hadrons that could potentially be discovered in future, and their names under the scheme set out in this document, are given in Table~\ref{tab:summary-prospective} with further discussion below.
\begin{itemize}
\item  All entries in  Table~\ref{tab:summary-prospective} are manifestly exotic, in that their minimal quark content does not contain $u\bar{u}$ or $d\bar{d}$.
States with such quark content could, however, potentially be identified as isospin partners of manifestly exotic states, in which scenario this naming scheme should also be applied to them.
\item The antiparticle of the hypothetical $c\bar{c}b\bar{d}$ state, with quark content $c\bar{c}\bar{b}d$, would be labelled as $\overline{T}{}^{\theta}_{\psi b1}({\rm mass})^0$.
\item The antiparticle of the hypothetical $c\bar{s}\bar{u}d$ state, with quark content $\bar{c}su\bar{d}$, would be labelled as $\overline{T}{}^{a}_{c\bar{s}0}({\rm mass})^0$.
\item The majority of the charged $T$ states would not require overlines to distinguish particle from antiparticle.  These would be required, however, for $I=1$ $T_{bc}$ or $T_{b\bar{s}}$ states, in the same way as in the example given above for the charged $T_{cs}$ states.  
\item % Particle and antiparticle for all charged $T$ states are distinguished by the charge only --- no overline is used in the symbol.  By contrast,
  All $P$ states with positive baryon number have symbols without, and all those with negative baryon number have symbols with, an overline (for example, the antiparticle of the $P_c(4312)^+$ is the $\overline{P}{}_c(4312)^-$ state).
\end{itemize}

\begin{table}[!tb]
  \renewcommand*{\arraystretch}{1.15}
  \centering
  \caption{
    Speculative examples of possible future discoveries of exotic hadrons.
    The hypothetical quantum numbers are in some cases based on theoretical expectations.
    Masses are indicated as $({\rm mass})$ and, for $P$ states, spin-parity labels are omitted.
    The potential decay channels quoted for the $bb\bar{u}\bar{d}$ state assume weak decays, as this state is expected to be below threshold for strong or electromagnetic decays~\cite{Ader:1981db,Manohar:1992nd,Bicudo:2015vta,Francis:2016hui,Karliner:2017qjm,Eichten:2017ffp,Czarnecki:2017vco}.
    All other potential decay channels assume strong decays.
  }
  \label{tab:summary-prospective}
  \resizebox{\textwidth}{!}{
  \begin{tabular}{cccc}
    \hline
    {Minimal quark} & \multirow{2}{*}{Potential decay channel(s)} & \multirow{2}{*}{$I^{(G)}$, $J^{P(C)}$} & \multirow{2}{*}{Proposed name} \\
    {content} \\
    \hline
    $bc\bar{u}\bar{d}$ & $\Bm\Dstarp$ & $I = 0$, $J^P = 1^+$ & $T^{f}_{bc1}({\rm mass})^0$ \\
    $b\bar{c}\bar{u}d$ & $\Bm\Dstarm$ & $I = 1$, $J^P = 1^+$ & $T^{a}_{b\bar{c}1}({\rm mass})^{--}$ \\
    $bb\bar{u}\bar{d}$ & $\Bm\pim\Dp$, $\Bzb\jpsi\Km$ & $I = 0$, $J^P = 1^+$ & $T^{f}_{bb1}({\rm mass})^-$ \\
    $c\bar{c}b\bar{d}$ & $\jpsi \Bzb$ & $I = \frac{1}{2}$, $J^P = 1^+$ & $T^{\theta}_{\psi b1}({\rm mass})^0$ \\
    $c\bar{s}u\bar{d}$/$c\bar{s}\bar{u}d$ & $\Dsp\pip$/$\Dsp\pim$ & $I = 1$, $J^P = 0^+$ & $T^{a}_{c\bar{s}0}({\rm mass})^{++}$/$T^{a}_{c\bar{s}0}({\rm mass})^0$ \\
    $b\bar{b}uud$ & $\Upsilonres\proton$ & $I = \frac{1}{2}$ & $P^{N}_{\Upsilonres}({\rm mass})^+$ \\
    $b\bar{c}uud$ & $\Bcm\proton$ & $I = \frac{1}{2}$ & $P^{N}_{b\bar{c}}({\rm mass})^0$ \\
    $b\bar{u}cds$ & $\Bm\Xicz$ & $I = 1$ & $P^{\PSigma}_{bcs}({\rm mass})^-$ \\
    $c\bar{d}cus$ & $\Dp\Xicp$ & $I = 1$ & $P^{\PSigma}_{ccs}({\rm mass})^{++}$ \\
    $c\bar{c}cud$ & $\jpsi\Lc$ & $I = 0$ & $P^{\PLambda}_{\psi c}({\rm mass})^+$ \\
    $c\bar{c}cus$ & $\jpsi\Xicp$ & $I = \frac{1}{2}$ & $P^{N}_{\psi cs}({\rm mass})^+$ \\
    \hline
  \end{tabular}
  }
\end{table}

\section{Summary}
\label{sec:summary}

A naming scheme for exotic hadrons, extending the existing conventions~\cite{RPP-naming}, has been set out.
It satisfies the goal of conveying in a reasonably compact way the quantum numbers of the states, and covers all currently known hadrons and many that could potentially be discovered in future.
It can also be extended further to include additional states that are not currently included in the scheme, such as six- or seven-quark hadrons.  
Due to the large number of quantum numbers that should be encoded in the name, the notation includes both superscripts and subscripts.
This may be inconvenient for typesetting on some systems, but it is anticipated that simplified names will be used for the most widely discussed states. 

This document covers the naming of hadrons, but it is also important to have a unique convention for a numbering scheme that can be used in Monte Carlo generators.
The current Monte Carlo particle numbering scheme~\cite{RPP-numbering} does not cover well the range of exotic hadrons that have recently been discovered.
It is hoped that a well-defined naming convention can also provide the basis for a numbering scheme.

\section*{Acknowledgements}

The development of this naming scheme for exotic hadrons has benefitted from input from many members of the community.
We would particularly like to thank the following for their constructive suggestions: Wolfgang Gradl, Christoph Hanhart, S\"{o}ren Lange, Ryan Mitchell, Eulogio Oset, Elisabetta Pianori, Elisabetta Prencipe, Ulrike Thoma, Diego Tonelli and Ron Workman.
We also appreciate invitations from the Belle~II, BESIII and PANDA collaborations to discuss the ideas in this document, and are grateful for the constructive feedback that was received at those meetings.

\addcontentsline{toc}{section}{References}
%\setboolean{inbibliography}{true}
\bibliographystyle{LHCb}
\bibliography{main,standard,LHCb-PAPER,LHCb-CONF,LHCb-DP,LHCb-TDR}

\ifx\mcitethebibliography\mciteundefinedmacro
\PackageError{LHCb.bst}{mciteplus.sty has not been loaded}
{This bibstyle requires the use of the mciteplus package.}\fi
\providecommand{\href}[2]{#2}
\begin{mcitethebibliography}{10}
\mciteSetBstSublistMode{n}
\mciteSetBstMaxWidthForm{subitem}{\alph{mcitesubitemcount})}
\mciteSetBstSublistLabelBeginEnd{\mcitemaxwidthsubitemform\space}
{\relax}{\relax}

\bibitem{LHCb-FIGURE-2021-001}
LHCb collaboration, \ifthenelse{\boolean{articletitles}}{\emph{{List of hadrons
  observed at the LHC}}, }{}
  \href{http://cdsweb.cern.ch/search?p=LHCb-FIGURE-2021-001&f=reportnumber&action_search=Search&c=LHCb+Figures}
  {LHCb-FIGURE-2021-001}, 2021, {Up-to-date list maintained at
  \href{https://www.nikhef.nl/~pkoppenb/particles.html}{\url{https://www.nikhef.nl/~pkoppenb/particles.html}}}\relax
\mciteBstWouldAddEndPuncttrue
\mciteSetBstMidEndSepPunct{\mcitedefaultmidpunct}
{\mcitedefaultendpunct}{\mcitedefaultseppunct}\relax
\EndOfBibitem
\bibitem{LHCb-PAPER-2015-029}
LHCb collaboration, R.~Aaij {\em et~al.},
  \ifthenelse{\boolean{articletitles}}{\emph{{Observation of $\jpsi\proton$
  resonances consistent with pentaquark states in
  \mbox{\decay{\Lb}{\jpsi\proton\Km}} decays}},
  }{}\href{https://doi.org/10.1103/PhysRevLett.115.072001}{Phys.\ Rev.\ Lett.\
  \textbf{115} (2015) 072001},
  \href{http://arxiv.org/abs/1507.03414}{{\normalfont\ttfamily
  arXiv:1507.03414}}\relax
\mciteBstWouldAddEndPuncttrue
\mciteSetBstMidEndSepPunct{\mcitedefaultmidpunct}
{\mcitedefaultendpunct}{\mcitedefaultseppunct}\relax
\EndOfBibitem
\bibitem{LHCb-PAPER-2019-014}
LHCb collaboration, R.~Aaij {\em et~al.},
  \ifthenelse{\boolean{articletitles}}{\emph{{Observation of a narrow
  pentaquark state, $P_c(4312)^+$, and of two-peak structure of the
  $P_c(4450)^+$}},
  }{}\href{https://doi.org/10.1103/PhysRevLett.122.222001}{Phys.\ Rev.\ Lett.\
  \textbf{122} (2019) 222001},
  \href{http://arxiv.org/abs/1904.03947}{{\normalfont\ttfamily
  arXiv:1904.03947}}\relax
\mciteBstWouldAddEndPuncttrue
\mciteSetBstMidEndSepPunct{\mcitedefaultmidpunct}
{\mcitedefaultendpunct}{\mcitedefaultseppunct}\relax
\EndOfBibitem
\bibitem{LHCb-PAPER-2020-011}
LHCb collaboration, R.~Aaij {\em et~al.},
  \ifthenelse{\boolean{articletitles}}{\emph{{Observation of structure in the
  \jpsi-pair mass spectrum}},
  }{}\href{https://doi.org/10.1016/j.scib.2020.08.032}{Science Bulletin
  \textbf{65} (2020) 1983},
  \href{http://arxiv.org/abs/2006.16957}{{\normalfont\ttfamily
  arXiv:2006.16957}}\relax
\mciteBstWouldAddEndPuncttrue
\mciteSetBstMidEndSepPunct{\mcitedefaultmidpunct}
{\mcitedefaultendpunct}{\mcitedefaultseppunct}\relax
\EndOfBibitem
\bibitem{LHCb-PAPER-2020-024}
LHCb collaboration, R.~Aaij {\em et~al.},
  \ifthenelse{\boolean{articletitles}}{\emph{{Model-independent study of
  structure in $\Bp \to \Dp \Dm \Kp$ decays}},
  }{}\href{https://doi.org/10.1103/PhysRevLett.125.242001}{Phys.\ Rev.\ Lett.\
  \textbf{125} (2020) 242001},
  \href{http://arxiv.org/abs/2009.00025}{{\normalfont\ttfamily
  arXiv:2009.00025}}\relax
\mciteBstWouldAddEndPuncttrue
\mciteSetBstMidEndSepPunct{\mcitedefaultmidpunct}
{\mcitedefaultendpunct}{\mcitedefaultseppunct}\relax
\EndOfBibitem
\bibitem{LHCb-PAPER-2020-025}
LHCb collaboration, R.~Aaij {\em et~al.},
  \ifthenelse{\boolean{articletitles}}{\emph{{Amplitude analysis of the $\Bp
  \to \Dp \Dm \Kp$ decay}},
  }{}\href{https://doi.org/10.1103/PhysRevD.102.112003}{Phys.\ Rev.\
  \textbf{D102} (2020) 112003},
  \href{http://arxiv.org/abs/2009.00026}{{\normalfont\ttfamily
  arXiv:2009.00026}}\relax
\mciteBstWouldAddEndPuncttrue
\mciteSetBstMidEndSepPunct{\mcitedefaultmidpunct}
{\mcitedefaultendpunct}{\mcitedefaultseppunct}\relax
\EndOfBibitem
\bibitem{LHCb-PAPER-2020-044}
LHCb collaboration, R.~Aaij {\em et~al.},
  \ifthenelse{\boolean{articletitles}}{\emph{{Observation of new resonances
  decaying to $ \jpsi K^+$ and $ \jpsi \phi$ }},
  }{}\href{https://doi.org/10.1103/PhysRevLett.127.082001}{Phys.\ Rev.\ Lett.\
  \textbf{127} (2021) 082001},
  \href{http://arxiv.org/abs/2103.01803}{{\normalfont\ttfamily
  arXiv:2103.01803}}\relax
\mciteBstWouldAddEndPuncttrue
\mciteSetBstMidEndSepPunct{\mcitedefaultmidpunct}
{\mcitedefaultendpunct}{\mcitedefaultseppunct}\relax
\EndOfBibitem
\bibitem{LHCb-PAPER-2021-031}
LHCb collaboration, R.~Aaij {\em et~al.},
  \ifthenelse{\boolean{articletitles}}{\emph{{Observation of an exotic narrow
  doubly charmed tetraquark}},
  }{}\href{https://doi.org/10.1038/s41567-022-01614-y}{Nature Physics
  \textbf{18} (2022) 751},
  \href{http://arxiv.org/abs/2109.01038}{{\normalfont\ttfamily
  arXiv:2109.01038}}\relax
\mciteBstWouldAddEndPuncttrue
\mciteSetBstMidEndSepPunct{\mcitedefaultmidpunct}
{\mcitedefaultendpunct}{\mcitedefaultseppunct}\relax
\EndOfBibitem
\bibitem{LHCb-PAPER-2021-032}
LHCb collaboration, R.~Aaij {\em et~al.},
  \ifthenelse{\boolean{articletitles}}{\emph{{Study of the doubly charmed
  tetraquark $T^+_{cc}$}},
  }{}\href{https://doi.org/10.1038/s41467-022-30206-w}{Nature Commun.\
  \textbf{13} (2022) 3351},
  \href{http://arxiv.org/abs/2109.01056}{{\normalfont\ttfamily
  arXiv:2109.01056}}\relax
\mciteBstWouldAddEndPuncttrue
\mciteSetBstMidEndSepPunct{\mcitedefaultmidpunct}
{\mcitedefaultendpunct}{\mcitedefaultseppunct}\relax
\EndOfBibitem
\bibitem{Ali:2017jda}
A.~Ali, J.~S. Lange, and S.~Stone,
  \ifthenelse{\boolean{articletitles}}{\emph{{Exotics: Heavy pentaquarks and
  tetraquarks}}, }{}\href{https://doi.org/10.1016/j.ppnp.2017.08.003}{Prog.\
  Part.\ Nucl.\ Phys.\  \textbf{97} (2017) 123},
  \href{http://arxiv.org/abs/1706.00610}{{\normalfont\ttfamily
  arXiv:1706.00610}}\relax
\mciteBstWouldAddEndPuncttrue
\mciteSetBstMidEndSepPunct{\mcitedefaultmidpunct}
{\mcitedefaultendpunct}{\mcitedefaultseppunct}\relax
\EndOfBibitem
\bibitem{Olsen:2017bmm}
S.~L. Olsen, T.~Skwarnicki, and D.~Zieminska,
  \ifthenelse{\boolean{articletitles}}{\emph{{Nonstandard heavy mesons and
  baryons: Experimental evidence}},
  }{}\href{https://doi.org/10.1103/RevModPhys.90.015003}{Rev.\ Mod.\ Phys.\
  \textbf{90} (2018) 015003},
  \href{http://arxiv.org/abs/1708.04012}{{\normalfont\ttfamily
  arXiv:1708.04012}}\relax
\mciteBstWouldAddEndPuncttrue
\mciteSetBstMidEndSepPunct{\mcitedefaultmidpunct}
{\mcitedefaultendpunct}{\mcitedefaultseppunct}\relax
\EndOfBibitem
\bibitem{Karliner:2017qhf}
M.~Karliner, J.~L. Rosner, and T.~Skwarnicki,
  \ifthenelse{\boolean{articletitles}}{\emph{{Multiquark states}},
  }{}\href{https://doi.org/10.1146/annurev-nucl-101917-020902}{Ann.\ Rev.\
  Nucl.\ Part.\ Sci.\  \textbf{68} (2018) 17},
  \href{http://arxiv.org/abs/1711.10626}{{\normalfont\ttfamily
  arXiv:1711.10626}}\relax
\mciteBstWouldAddEndPuncttrue
\mciteSetBstMidEndSepPunct{\mcitedefaultmidpunct}
{\mcitedefaultendpunct}{\mcitedefaultseppunct}\relax
\EndOfBibitem
\bibitem{Guo:2017jvc}
F.-K. Guo {\em et~al.}, \ifthenelse{\boolean{articletitles}}{\emph{{Hadronic
  molecules}}, }{}\href{https://doi.org/10.1103/RevModPhys.90.015004}{Rev.\
  Mod.\ Phys.\  \textbf{90} (2018) 015004},
  \href{http://arxiv.org/abs/1705.00141}{{\normalfont\ttfamily
  arXiv:1705.00141}}\relax
\mciteBstWouldAddEndPuncttrue
\mciteSetBstMidEndSepPunct{\mcitedefaultmidpunct}
{\mcitedefaultendpunct}{\mcitedefaultseppunct}\relax
\EndOfBibitem
\bibitem{Liu:2019zoy}
Y.-R. Liu {\em et~al.}, \ifthenelse{\boolean{articletitles}}{\emph{{Pentaquark
  and tetraquark states}},
  }{}\href{https://doi.org/10.1016/j.ppnp.2019.04.003}{Prog.\ Part.\ Nucl.\
  Phys.\  \textbf{107} (2019) 237},
  \href{http://arxiv.org/abs/1903.11976}{{\normalfont\ttfamily
  arXiv:1903.11976}}\relax
\mciteBstWouldAddEndPuncttrue
\mciteSetBstMidEndSepPunct{\mcitedefaultmidpunct}
{\mcitedefaultendpunct}{\mcitedefaultseppunct}\relax
\EndOfBibitem
\bibitem{Brambilla:2019esw}
N.~Brambilla {\em et~al.}, \ifthenelse{\boolean{articletitles}}{\emph{{The
  $XYZ$ states: Experimental and theoretical status and perspectives}},
  }{}\href{https://doi.org/10.1016/j.physrep.2020.05.001}{Phys.\ Rept.\
  \textbf{873} (2020) 1},
  \href{http://arxiv.org/abs/1907.07583}{{\normalfont\ttfamily
  arXiv:1907.07583}}\relax
\mciteBstWouldAddEndPuncttrue
\mciteSetBstMidEndSepPunct{\mcitedefaultmidpunct}
{\mcitedefaultendpunct}{\mcitedefaultseppunct}\relax
\EndOfBibitem
\bibitem{PDG2020}
Particle Data Group, P.~A. Zyla {\em et~al.},
  \ifthenelse{\boolean{articletitles}}{\emph{{\href{http://pdg.lbl.gov/}{Review
  of particle physics}}}, }{}\href{https://doi.org/10.1093/ptep/ptaa104}{Prog.\
  Theor.\ Exp.\ Phys.\  \textbf{2020} (2020) 083C01}\relax
\mciteBstWouldAddEndPuncttrue
\mciteSetBstMidEndSepPunct{\mcitedefaultmidpunct}
{\mcitedefaultendpunct}{\mcitedefaultseppunct}\relax
\EndOfBibitem
\bibitem{CERNCourierVolume25:1731195}
M.~Roos, \ifthenelse{\boolean{articletitles}}{\emph{{New names for old
  mesons}}, }{}{CERN Courier} \textbf{25} (1985) 390, {Available at
  \href{https://cds.cern.ch/record/1731193}{\url{https://cds.cern.ch/record/1731193}}}\relax
\mciteBstWouldAddEndPuncttrue
\mciteSetBstMidEndSepPunct{\mcitedefaultmidpunct}
{\mcitedefaultendpunct}{\mcitedefaultseppunct}\relax
\EndOfBibitem
\bibitem{RPP-naming}
V.~Burkert {\em et~al.}, \ifthenelse{\boolean{articletitles}}{\emph{{Naming
  scheme for hadrons}}, }{}
\newblock {in Ref.~\cite{PDG2020}}\relax
\mciteBstWouldAddEndPuncttrue
\mciteSetBstMidEndSepPunct{\mcitedefaultmidpunct}
{\mcitedefaultendpunct}{\mcitedefaultseppunct}\relax
\EndOfBibitem
\bibitem{RPP-quarkmodel}
C.~Amsler, T.~DeGrand, and B.~Krusche,
  \ifthenelse{\boolean{articletitles}}{\emph{{Quark model}}, }{}
\newblock {in Ref.~\cite{PDG2020}}\relax
\mciteBstWouldAddEndPuncttrue
\mciteSetBstMidEndSepPunct{\mcitedefaultmidpunct}
{\mcitedefaultendpunct}{\mcitedefaultseppunct}\relax
\EndOfBibitem
\bibitem{LHCb-PAPER-2020-039}
LHCb collaboration, R.~Aaij {\em et~al.},
  \ifthenelse{\boolean{articletitles}}{\emph{{Evidence of a $\jpsi\Lz$
  structure and observation of excited $\Xim$ states in the $\Xibm \rightarrow
  \jpsi \Lz \Km$ decay}},
  }{}\href{https://doi.org/10.1016/j.scib.2021.02.030}{Science Bulletin
  \textbf{66} (2021) 1278},
  \href{http://arxiv.org/abs/2012.10380}{{\normalfont\ttfamily
  arXiv:2012.10380}}\relax
\mciteBstWouldAddEndPuncttrue
\mciteSetBstMidEndSepPunct{\mcitedefaultmidpunct}
{\mcitedefaultendpunct}{\mcitedefaultseppunct}\relax
\EndOfBibitem
\bibitem{Dalitz:1953cp}
R.~H. Dalitz, \ifthenelse{\boolean{articletitles}}{\emph{{On the analysis of
  \Ptau-meson data and the nature of the \Ptau-meson}},
  }{}\href{https://doi.org/10.1080/14786441008520365}{Phil.\ Mag.\ Ser.\ 7
  \textbf{44} (1953) 1068}\relax
\mciteBstWouldAddEndPuncttrue
\mciteSetBstMidEndSepPunct{\mcitedefaultmidpunct}
{\mcitedefaultendpunct}{\mcitedefaultseppunct}\relax
\EndOfBibitem
\bibitem{Lee:1956qn}
T.~D. Lee and C.-N. Yang, \ifthenelse{\boolean{articletitles}}{\emph{{Question
  of parity conservation in weak interactions}},
  }{}\href{https://doi.org/10.1103/PhysRev.104.254}{Phys.\ Rev.\  \textbf{104}
  (1956) 254}\relax
\mciteBstWouldAddEndPuncttrue
\mciteSetBstMidEndSepPunct{\mcitedefaultmidpunct}
{\mcitedefaultendpunct}{\mcitedefaultseppunct}\relax
\EndOfBibitem
\bibitem{Dalitz:1993yt}
R.~H. Dalitz, \ifthenelse{\boolean{articletitles}}{\emph{{The $\tau-\theta$
  puzzle}}, }{}\href{https://doi.org/10.1063/1.45424}{AIP Conf.\ Proc.\
  \textbf{300} (1994) 141}\relax
\mciteBstWouldAddEndPuncttrue
\mciteSetBstMidEndSepPunct{\mcitedefaultmidpunct}
{\mcitedefaultendpunct}{\mcitedefaultseppunct}\relax
\EndOfBibitem
\bibitem{Belle:2003nnu}
Belle collaboration, S.~K. Choi {\em et~al.},
  \ifthenelse{\boolean{articletitles}}{\emph{{Observation of a narrow
  charmoniumlike state in exclusive $B^\pm \to K^\pm \pi^+ \pi^- J/\psi$
  decays}}, }{}\href{https://doi.org/10.1103/PhysRevLett.91.262001}{Phys.\
  Rev.\ Lett.\  \textbf{91} (2003) 262001},
  \href{http://arxiv.org/abs/hep-ex/0309032}{{\normalfont\ttfamily
  arXiv:hep-ex/0309032}}\relax
\mciteBstWouldAddEndPuncttrue
\mciteSetBstMidEndSepPunct{\mcitedefaultmidpunct}
{\mcitedefaultendpunct}{\mcitedefaultseppunct}\relax
\EndOfBibitem
\bibitem{LHCb-PAPER-2013-001}
LHCb collaboration, R.~Aaij {\em et~al.},
  \ifthenelse{\boolean{articletitles}}{\emph{{Determination of the $X(3872)$
  meson quantum numbers}},
  }{}\href{https://doi.org/10.1103/PhysRevLett.110.222001}{Phys.\ Rev.\ Lett.\
  \textbf{110} (2013) 222001},
  \href{http://arxiv.org/abs/1302.6269}{{\normalfont\ttfamily
  arXiv:1302.6269}}\relax
\mciteBstWouldAddEndPuncttrue
\mciteSetBstMidEndSepPunct{\mcitedefaultmidpunct}
{\mcitedefaultendpunct}{\mcitedefaultseppunct}\relax
\EndOfBibitem
\bibitem{BESIII:2013ris}
BESIII collaboration, M.~Ablikim {\em et~al.},
  \ifthenelse{\boolean{articletitles}}{\emph{{Observation of a charged
  charmoniumlike structure in $e^+e^- \to \pi^+\pi^- \jpsi$ at $\sqrt{s} =4.26$
  GeV}}, }{}\href{https://doi.org/10.1103/PhysRevLett.110.252001}{Phys.\ Rev.\
  Lett.\  \textbf{110} (2013) 252001},
  \href{http://arxiv.org/abs/1303.5949}{{\normalfont\ttfamily
  arXiv:1303.5949}}\relax
\mciteBstWouldAddEndPuncttrue
\mciteSetBstMidEndSepPunct{\mcitedefaultmidpunct}
{\mcitedefaultendpunct}{\mcitedefaultseppunct}\relax
\EndOfBibitem
\bibitem{Belle:2013yex}
Belle collaboration, Z.~Q. Liu {\em et~al.},
  \ifthenelse{\boolean{articletitles}}{\emph{{Study of $e^+e^- \to \pi^+ \pi^-
  \jpsi$ and observation of a charged charmoniumlike state at Belle}},
  }{}\href{https://doi.org/10.1103/PhysRevLett.110.252002}{Phys.\ Rev.\ Lett.\
  \textbf{110} (2013) 252002}, Erratum
  \href{https://doi.org/10.1103/PhysRevLett.111.019901}{ibid.\   \textbf{111}
  (2013) 019901}, \href{http://arxiv.org/abs/1304.0121}{{\normalfont\ttfamily
  arXiv:1304.0121}}\relax
\mciteBstWouldAddEndPuncttrue
\mciteSetBstMidEndSepPunct{\mcitedefaultmidpunct}
{\mcitedefaultendpunct}{\mcitedefaultseppunct}\relax
\EndOfBibitem
\bibitem{BESIII:2017bua}
BESIII collaboration, M.~Ablikim {\em et~al.},
  \ifthenelse{\boolean{articletitles}}{\emph{{Determination of the spin and
  parity of the $Z_c(3900)$}},
  }{}\href{https://doi.org/10.1103/PhysRevLett.119.072001}{Phys.\ Rev.\ Lett.\
  \textbf{119} (2017) 072001},
  \href{http://arxiv.org/abs/1706.04100}{{\normalfont\ttfamily
  arXiv:1706.04100}}\relax
\mciteBstWouldAddEndPuncttrue
\mciteSetBstMidEndSepPunct{\mcitedefaultmidpunct}
{\mcitedefaultendpunct}{\mcitedefaultseppunct}\relax
\EndOfBibitem
\bibitem{LHCb-PAPER-2018-034}
LHCb collaboration, R.~Aaij {\em et~al.},
  \ifthenelse{\boolean{articletitles}}{\emph{{Evidence for a $\eta_c(1S) \pim$
  resonance in \mbox{\decay{\Bz}{\eta_c(1S) \Kp\pim}} decays}},
  }{}\href{https://doi.org/10.1140/epjc/s10052-018-6447-z}{Eur.\ Phys.\ J.\
  \textbf{C78} (2018) 1019},
  \href{http://arxiv.org/abs/1809.07416}{{\normalfont\ttfamily
  arXiv:1809.07416}}\relax
\mciteBstWouldAddEndPuncttrue
\mciteSetBstMidEndSepPunct{\mcitedefaultmidpunct}
{\mcitedefaultendpunct}{\mcitedefaultseppunct}\relax
\EndOfBibitem
\bibitem{Belle:2007hrb}
Belle collaboration, S.~K. Choi {\em et~al.},
  \ifthenelse{\boolean{articletitles}}{\emph{{Observation of a resonancelike
  structure in the $\pi^\pm \psi^\prime$ mass distribution in exclusive $B \to
  K \pi^\pm \psi^\prime$ decays}},
  }{}\href{https://doi.org/10.1103/PhysRevLett.100.142001}{Phys.\ Rev.\ Lett.\
  \textbf{100} (2008) 142001},
  \href{http://arxiv.org/abs/0708.1790}{{\normalfont\ttfamily
  arXiv:0708.1790}}\relax
\mciteBstWouldAddEndPuncttrue
\mciteSetBstMidEndSepPunct{\mcitedefaultmidpunct}
{\mcitedefaultendpunct}{\mcitedefaultseppunct}\relax
\EndOfBibitem
\bibitem{LHCb-PAPER-2014-014}
LHCb collaboration, R.~Aaij {\em et~al.},
  \ifthenelse{\boolean{articletitles}}{\emph{{Observation of the resonant
  character of the $Z(4430)^-$ state}},
  }{}\href{https://doi.org/10.1103/PhysRevLett.112.222002}{Phys.\ Rev.\ Lett.\
  \textbf{112} (2014) 222002},
  \href{http://arxiv.org/abs/1404.1903}{{\normalfont\ttfamily
  arXiv:1404.1903}}\relax
\mciteBstWouldAddEndPuncttrue
\mciteSetBstMidEndSepPunct{\mcitedefaultmidpunct}
{\mcitedefaultendpunct}{\mcitedefaultseppunct}\relax
\EndOfBibitem
\bibitem{CDF:2009jgo}
CDF collaboration, T.~Aaltonen {\em et~al.},
  \ifthenelse{\boolean{articletitles}}{\emph{{Evidence for a narrow
  near-threshold structure in the $\jpsi\phi$ mass spectrum in $\Bp\to
  \jpsi\phi \Kp$ decays}},
  }{}\href{https://doi.org/10.1103/PhysRevLett.102.242002}{Phys.\ Rev.\ Lett.\
  \textbf{102} (2009) 242002},
  \href{http://arxiv.org/abs/0903.2229}{{\normalfont\ttfamily
  arXiv:0903.2229}}\relax
\mciteBstWouldAddEndPuncttrue
\mciteSetBstMidEndSepPunct{\mcitedefaultmidpunct}
{\mcitedefaultendpunct}{\mcitedefaultseppunct}\relax
\EndOfBibitem
\bibitem{CMS:2013jru}
CMS collaboration, S.~Chatrchyan {\em et~al.},
  \ifthenelse{\boolean{articletitles}}{\emph{{Observation of a peaking
  structure in the $\jpsi \phi$ mass spectrum from $\Bpm \to \jpsi \phi \Kpm$
  decays}}, }{}\href{https://doi.org/10.1016/j.physletb.2014.05.055}{Phys.\
  Lett.\  \textbf{B734} (2014) 261},
  \href{http://arxiv.org/abs/1309.6920}{{\normalfont\ttfamily
  arXiv:1309.6920}}\relax
\mciteBstWouldAddEndPuncttrue
\mciteSetBstMidEndSepPunct{\mcitedefaultmidpunct}
{\mcitedefaultendpunct}{\mcitedefaultseppunct}\relax
\EndOfBibitem
\bibitem{LHCb-PAPER-2016-018}
LHCb collaboration, R.~Aaij {\em et~al.},
  \ifthenelse{\boolean{articletitles}}{\emph{{Observation of exotic $\jpsi\phi$
  structures from amplitude analysis of \mbox{\decay{\Bp}{\jpsi\phi\Kp}}
  decays}}, }{}\href{https://doi.org/10.1103/PhysRevLett.118.022003}{Phys.\
  Rev.\ Lett.\  \textbf{118} (2017) 022003},
  \href{http://arxiv.org/abs/1606.07895}{{\normalfont\ttfamily
  arXiv:1606.07895}}\relax
\mciteBstWouldAddEndPuncttrue
\mciteSetBstMidEndSepPunct{\mcitedefaultmidpunct}
{\mcitedefaultendpunct}{\mcitedefaultseppunct}\relax
\EndOfBibitem
\bibitem{LHCb-PAPER-2016-019}
LHCb collaboration, R.~Aaij {\em et~al.},
  \ifthenelse{\boolean{articletitles}}{\emph{{Amplitude analysis of
  \mbox{\decay{\Bp}{\jpsi\phi\Kp}} decays}},
  }{}\href{https://doi.org/10.1103/PhysRevD.95.012002}{Phys.\ Rev.\
  \textbf{D95} (2017) 012002},
  \href{http://arxiv.org/abs/1606.07898}{{\normalfont\ttfamily
  arXiv:1606.07898}}\relax
\mciteBstWouldAddEndPuncttrue
\mciteSetBstMidEndSepPunct{\mcitedefaultmidpunct}
{\mcitedefaultendpunct}{\mcitedefaultseppunct}\relax
\EndOfBibitem
\bibitem{Belle:2011aa}
Belle collaboration, A.~Bondar {\em et~al.},
  \ifthenelse{\boolean{articletitles}}{\emph{{Observation of two charged
  bottomoniumlike resonances in $\Upsilonres(5S)$ decays}},
  }{}\href{https://doi.org/10.1103/PhysRevLett.108.122001}{Phys.\ Rev.\ Lett.\
  \textbf{108} (2012) 122001},
  \href{http://arxiv.org/abs/1110.2251}{{\normalfont\ttfamily
  arXiv:1110.2251}}\relax
\mciteBstWouldAddEndPuncttrue
\mciteSetBstMidEndSepPunct{\mcitedefaultmidpunct}
{\mcitedefaultendpunct}{\mcitedefaultseppunct}\relax
\EndOfBibitem
\bibitem{BaBar:2003oey}
BaBar collaboration, B.~Aubert {\em et~al.},
  \ifthenelse{\boolean{articletitles}}{\emph{{Observation of a narrow meson
  decaying to $D_s^+ \pi^0$ at a mass of $2.32 \gevcc$}},
  }{}\href{https://doi.org/10.1103/PhysRevLett.90.242001}{Phys.\ Rev.\ Lett.\
  \textbf{90} (2003) 242001},
  \href{http://arxiv.org/abs/hep-ex/0304021}{{\normalfont\ttfamily
  arXiv:hep-ex/0304021}}\relax
\mciteBstWouldAddEndPuncttrue
\mciteSetBstMidEndSepPunct{\mcitedefaultmidpunct}
{\mcitedefaultendpunct}{\mcitedefaultseppunct}\relax
\EndOfBibitem
\bibitem{CLEO:2003ggt}
CLEO collaboration, D.~Besson {\em et~al.},
  \ifthenelse{\boolean{articletitles}}{\emph{{Observation of a narrow resonance
  of mass $2.46 \gevcc$ decaying to $D^{*+}_s\pi^0$ and confirmation of the
  $D^*_{sJ}(2317)$ state}},
  }{}\href{https://doi.org/10.1103/PhysRevD.68.032002}{Phys.\ Rev.\
  \textbf{D68} (2003) 032002}, Erratum
  \href{https://doi.org/10.1103/PhysRevD.75.119908}{ibid.\   \textbf{D75}
  (2007) 119908},
  \href{http://arxiv.org/abs/hep-ex/0305100}{{\normalfont\ttfamily
  arXiv:hep-ex/0305100}}\relax
\mciteBstWouldAddEndPuncttrue
\mciteSetBstMidEndSepPunct{\mcitedefaultmidpunct}
{\mcitedefaultendpunct}{\mcitedefaultseppunct}\relax
\EndOfBibitem
\bibitem{Ader:1981db}
J.-P. Ader, J.-M. Richard, and P.~Taxil,
  \ifthenelse{\boolean{articletitles}}{\emph{{Do narrow heavy multi-quark
  states exist?}}, }{}\href{https://doi.org/10.1103/PhysRevD.25.2370}{Phys.\
  Rev.\  \textbf{D25} (1982) 2370}\relax
\mciteBstWouldAddEndPuncttrue
\mciteSetBstMidEndSepPunct{\mcitedefaultmidpunct}
{\mcitedefaultendpunct}{\mcitedefaultseppunct}\relax
\EndOfBibitem
\bibitem{Manohar:1992nd}
A.~V. Manohar and M.~B. Wise,
  \ifthenelse{\boolean{articletitles}}{\emph{{Exotic $QQ \bar{q}\bar{q}$ states
  in QCD}}, }{}\href{https://doi.org/10.1016/0550-3213(93)90614-U}{Nucl.\
  Phys.\  \textbf{B399} (1993) 17},
  \href{http://arxiv.org/abs/hep-ph/9212236}{{\normalfont\ttfamily
  arXiv:hep-ph/9212236}}\relax
\mciteBstWouldAddEndPuncttrue
\mciteSetBstMidEndSepPunct{\mcitedefaultmidpunct}
{\mcitedefaultendpunct}{\mcitedefaultseppunct}\relax
\EndOfBibitem
\bibitem{Bicudo:2015vta}
P.~Bicudo {\em et~al.}, \ifthenelse{\boolean{articletitles}}{\emph{{Evidence
  for the existence of $u d \bar{b} \bar{b}$ and the nonexistence of $s s
  \bar{b} \bar{b}$ and $c c \bar{b} \bar{b}$ tetraquarks from lattice QCD}},
  }{}\href{https://doi.org/10.1103/PhysRevD.92.014507}{Phys.\ Rev.\
  \textbf{D92} (2015) 014507},
  \href{http://arxiv.org/abs/1505.00613}{{\normalfont\ttfamily
  arXiv:1505.00613}}\relax
\mciteBstWouldAddEndPuncttrue
\mciteSetBstMidEndSepPunct{\mcitedefaultmidpunct}
{\mcitedefaultendpunct}{\mcitedefaultseppunct}\relax
\EndOfBibitem
\bibitem{Francis:2016hui}
A.~Francis, R.~J. Hudspith, R.~Lewis, and K.~Maltman,
  \ifthenelse{\boolean{articletitles}}{\emph{{Lattice prediction for deeply
  bound doubly heavy tetraquarks}},
  }{}\href{https://doi.org/10.1103/PhysRevLett.118.142001}{Phys.\ Rev.\ Lett.\
  \textbf{118} (2017) 142001},
  \href{http://arxiv.org/abs/1607.05214}{{\normalfont\ttfamily
  arXiv:1607.05214}}\relax
\mciteBstWouldAddEndPuncttrue
\mciteSetBstMidEndSepPunct{\mcitedefaultmidpunct}
{\mcitedefaultendpunct}{\mcitedefaultseppunct}\relax
\EndOfBibitem
\bibitem{Karliner:2017qjm}
M.~Karliner and J.~L. Rosner,
  \ifthenelse{\boolean{articletitles}}{\emph{{Discovery of doubly charmed
  $\Xires_{cc}$ baryon implies a stable ($b b \bar{u} \bar{d}$) tetraquark}},
  }{}\href{https://doi.org/10.1103/PhysRevLett.119.202001}{Phys.\ Rev.\ Lett.\
  \textbf{119} (2017) 202001},
  \href{http://arxiv.org/abs/1707.07666}{{\normalfont\ttfamily
  arXiv:1707.07666}}\relax
\mciteBstWouldAddEndPuncttrue
\mciteSetBstMidEndSepPunct{\mcitedefaultmidpunct}
{\mcitedefaultendpunct}{\mcitedefaultseppunct}\relax
\EndOfBibitem
\bibitem{Eichten:2017ffp}
E.~J. Eichten and C.~Quigg,
  \ifthenelse{\boolean{articletitles}}{\emph{{Heavy-quark symmetry implies
  stable heavy tetraquark mesons $Q_iQ_j \bar q_k \bar q_l$}},
  }{}\href{https://doi.org/10.1103/PhysRevLett.119.202002}{Phys.\ Rev.\ Lett.\
  \textbf{119} (2017) 202002},
  \href{http://arxiv.org/abs/1707.09575}{{\normalfont\ttfamily
  arXiv:1707.09575}}\relax
\mciteBstWouldAddEndPuncttrue
\mciteSetBstMidEndSepPunct{\mcitedefaultmidpunct}
{\mcitedefaultendpunct}{\mcitedefaultseppunct}\relax
\EndOfBibitem
\bibitem{Czarnecki:2017vco}
A.~Czarnecki, B.~Leng, and M.~B. Voloshin,
  \ifthenelse{\boolean{articletitles}}{\emph{{Stability of tetrons}},
  }{}\href{https://doi.org/10.1016/j.physletb.2018.01.034}{Phys.\ Lett.\
  \textbf{B778} (2018) 233},
  \href{http://arxiv.org/abs/1708.04594}{{\normalfont\ttfamily
  arXiv:1708.04594}}\relax
\mciteBstWouldAddEndPuncttrue
\mciteSetBstMidEndSepPunct{\mcitedefaultmidpunct}
{\mcitedefaultendpunct}{\mcitedefaultseppunct}\relax
\EndOfBibitem
\bibitem{RPP-numbering}
F.~Krauss, S.~Navas, P.~Richardson, and T.~Sj{\"o}strand,
  \ifthenelse{\boolean{articletitles}}{\emph{{Monte Carlo particle numbering
  scheme}}, }{}
\newblock {in Ref.~\cite{PDG2020}}\relax
\mciteBstWouldAddEndPuncttrue
\mciteSetBstMidEndSepPunct{\mcitedefaultmidpunct}
{\mcitedefaultendpunct}{\mcitedefaultseppunct}\relax
\EndOfBibitem
\end{mcitethebibliography}

\end{document}